\documentclass[12pt,preprint]{aastex}
\newcommand\Mj{\mbox{$M_{\rm Jup}$}}

\newcommand\Msun{\mbox{$M_\sun$}}
\newcommand\Lsun{\mbox{$L_\sun$}}

\shortauthors{McElwain et al.}
\shorttitle{IFS Spectroscopy of GQ Lup}

\begin{document}

\title{First High-Contrast Science with an Integral Field Spectrograph: the Sub-Stellar Companion to GQ Lup}

\author{Michael W.\ McElwain\altaffilmark{1}, Stanimir A.\
Metchev\altaffilmark{1,2}, James E.\ Larkin\altaffilmark{1}, Matthew
Barczys\altaffilmark{1}, Christof Iserlohe\altaffilmark{3}, Alfred
Krabbe\altaffilmark{3}, Andreas Quirrenbach\altaffilmark{4}, Jason
Weiss\altaffilmark{1}, Shelley A.\ Wright\altaffilmark{1}, }

\altaffiltext{1}{Department of Physics \& Astronomy, University of
California at Los Angeles, Los Angeles, CA, 90095-1562;
mcelwain@astro.ucla.edu, metchev@astro.ucla.edu,
larkin@astro.ucla.edu, barczysm@astro.ucla.edu, weiss@astro.ucla.edu,
saw@astro.ucla.edu}
\altaffiltext{2}{Spitzer Fellow}
\altaffiltext{3}{I. Physikalisches Institut, Universit\"at zu K\"oln,
50937 K\"oln, Germany; krabbe@ph1.uni-koeln.de, iserlohe@ph1.uni- 
koeln.de}
\altaffiltext{4}{ZAH Landessternwarte, Koenigstuhl, 
D-69117 Heidelberg, Germany; A.Quirrenbach@lsw.uni-heidelberg.de}

\received{}
\revised{}
\accepted{}
\journalid{}{}
\articleid{}{}

\begin{abstract}
We present commissioning data from the OSIRIS integral field
spectrograph (IFS) on the Keck~II 10~m telescope that demonstrate the
utility of adaptive optics IFS spectroscopy in studying faint close-in
sub-stellar companions in the haloes of bright stars.  Our
$R$$\approx$2000 $J$- and $H$-band spectra of the sub-stellar
companion to the 1--10~Myr-old GQ~Lup complement existing $K$-band
spectra and photometry, and improve on the original estimate of its
spectral type.  We find that GQ~Lup~B is somewhat hotter (M6--L0) than
reported in the discovery paper by Neuh{\" a}user and
collaborators (M9--L4), mainly due to the surface-gravity sensitivity of the
$K$-band spectral classification indices used by the discoverers.
Spectroscopic features characteristic of low surface gravity objects,
such as lack of alkali absorption and a triangular $H$-band continuum,
are indeed prominent in our spectrum of GQ~Lup~B.  The peculiar shape
of the $H$-band continuum and the difference between the two spectral
type estimates is well explained in the context of the diminishing
strength of H$_2$ collision induced absorption with decreasing surface
gravity, as recently proposed for young ultra-cool dwarfs by
Kirkpatrick and collaborators.  Using our updated spectroscopic
classification of GQ~Lup~B and a re-evaluation of the age and
heliocentric distance of the primary, we perform a comparative
analysis of the available sub-stellar evolutionary models to estimate
the mass of the companion.  We find that the mass of GQ~Lup~B is
0.010--0.040~\Msun.  Hence, it is unlikely to be a wide-orbit counterpart
to the known radial-velocity extrasolar planets, whose masses are
$\lesssim$0.015~\Msun.  Instead, GQ~Lup~A/B is probably a member of a
growing family of very low mass ratio widely separated binaries
discovered through high-contrast imaging.

\end{abstract}


\section{INTRODUCTION}

After more than a decade of precision radial velocity surveys, we know
that extrasolar giant planets exist around at least 5--15\% of
Sun-like stars \citep{marcy_butler00, fischer_etal03}.  Unfortunately,
these planets lie at small angular separations ($<$$0\farcs5$) with
high contrast ($>$$10^6$) from their host stars, and therefore, every
known radial-velocity extrasolar planet is beyond the current
technical limitations for direct imaging.  As a result, the physical
properties of these radial velocity planets (except for several
transiting planets) remain largely unknown.  However, recent efforts in
high-contrast imaging with adaptive optics (AO) have lead to the
discovery of two distinct sub-stellar companions,
2MASSW~J1207334--393254~B \citep{chauvin_etal05a} and GQ~Lup~B
\citep{neuhauser_etal05}, at wider ($>$$0\farcs7$) angular separations
from their primary stars and with estimated masses comparable to those
of known radial velocity planets ($\lesssim$15 Jupiter masses [\Mj]).
Unlike the close-in ($\lesssim$6~AU) radial velocity planets, whose
(minimum) masses are inferred directly from the orbital periodicity of
the Doppler signal, the masses of these two wider ($>$40~AU)
companions have not been established dynamically due to their much
longer orbital periods.  Instead, the estimated masses are based
entirely on theoretical models of sub-stellar evolution
\citep{burrows_etal97, burrows_etal01, chabrier_etal00,
baraffe_etal03, wuchterl_tscharnuter03}.  Presently, such models have
very few empirical constraints, and at the young ($\lesssim$10~Myr) ages of
the two directly-imaged planetary-mass companions, their predictions
are very sensitive to the initial conditions.  A larger sample of
empirical data on such young low-mass objects is thus necessary to
calibrate and fully understand the evolution of these objects at such
young ages.

In the present paper we discuss the younger of the two resolved
candidate planetary-mass companions, GQ~Lup~B.  The secondary was
discovered and confirmed as a proper motion companion to GQ~Lup~A by
\citet{neuhauser_etal05}.  Being one of the youngest low-mass
sub-stellar objects discovered to date, there is considerable
uncertainty in estimating the mass of GQ~Lup~B.
\citeauthor{neuhauser_etal05} argue that the widely-adopted
sub-stellar evolutionary models of \citet{burrows_etal97} and
\citet{chabrier_etal00}, which assume an initial post-formation
internal structure for sub-stellar objects, are inadequate at the
young age of GQ~Lup because of the arbitrariness in their assumptions
for the initial conditions.  Instead, \citeauthor{neuhauser_etal05}
advocate the use of a different set of models
\citep{wuchterl_tscharnuter03}, which take into account the conditions
and processes in the parent molecular cloud, and may be more adequate
at very young ($\sim$1~Myr) ages.  Using the models of
\citeauthor{wuchterl_tscharnuter03} and assuming an age of $\sim$1.1 Myr,
\citeauthor{neuhauser_etal05} find that the mass of GQ~Lup~B may be as
low as 1--2~\Mj, and hence that it is potentially the first extrasolar
planet to be directly imaged around a star.

We present new near-IR spectroscopic data on GQ Lup B obtained with
the recently commissioned OSIRIS (OH Suppressing Infra-Red Imaging
Spectrograph) IFS \citep{larkin_etal06} on Keck
(\S\ref{sec_observations}--\ref{sec_results}), recalculate the age of
GQ Lup, and address recent empirical calibrations of the sub-stellar
models (\S\ref{sec_analysis}).  We find that the mass of GQ~Lup~B is
near, or more likely higher than, 13~\Mj\ and thus should not be
considered as an extrasolar planet.  Finally, we place GQ~Lup~B in the
context of other young sub-stellar objects, and discuss the
implications from its co-evolution with the primordial circumstellar
disk around GQ~Lup (\S\ref{sec_disk}).

\section{OBSERVATIONS \label{sec_observations}}

\subsection{Integral Field Spectroscopy}

GQ Lup was observed on 2005 June 26 (UT) as a commissioning target for
the OSIRIS instrument at the W.M.~Keck Observatory.  OSIRIS is a
medium resolution ($R$=3700) infrared (IR) integral field spectrograph
(IFS) that was designed and constructed to operate behind the Keck
Observatory's AO System \citep{wizinowich_etal00}.  An IFS is an
instrument that takes contiguous spectra over a rectangular field of
view (FOV).  In the case of OSIRIS, a microlens array is placed in the
focal plane of the instrument to separate the field points, where each
lenslet becomes a spatial pixel element (spaxel) in the final data
cube.  Each lenslet focuses the incident light to a pupil plane
located directly behind the lenslet array.  These pupils are dispersed
by a diffraction grating and subsequently focused onto a detector in
such a way that each spectrum lies 2 pixels above its neighbor on the
detector.  The OSIRIS IFS employs a Rockwell Hawaii II HgCdTe detector
(2048$\times$2048 pixels, 32 channel output) to achieve high quantum
efficiency in the $z$, $J$, $H$, $K$ bands with low read noise (13 -e
pix$^{-1}$).  The actual near-IR filters are spectroscopic filters and
do not correspond exactly to any near-IR standard.  These broad bands
correspond to different orders of the diffraction grating, as defined
by the blaze angle.  \citet{larkin_etal06} presents a comprehensive
review of the OSIRIS instrument design.

We made natural guide star (NGS) AO observations of GQ Lup in the $J$
and $H$-broad bands over a 0$\farcs$32$\times$1$\farcs$28 FOV at
0$\farcs$020 spaxel$^{-1}$.  In each of the observations, the position
angle of the long axis FOV was set to 90$\degr$ in order to align the
rectagular field of view along the separation axis of the GQ~Lup~A and
B components.  The $J$-band images were obtained in a 3 point raster
scan pattern with a 0.15$\arcsec$ dither to the East between each of
the 600~s integrations, thus moving GQ~Lup~A from the center of the
field to just off the field of view in the final frame.  The science
frames were followed by a sky of similar integration time.  A single
600~s $H$-band image was obtained of the GQ~Lup system, with a
3$\arcsec$ dither to a sky position.  Observing conditions were
photometric, with the Keck AO system running on-axis using GQ~Lup~A
($R$=11~mag) as the reference star.  We measure the AO point spread
function (PSF) full-widths at half-maximum of GQ~Lup~A to be
0$\farcs$06 and 0$\farcs$05 at $J$ and $H$ band, respectively.
Afterwards, we observed the A0V star HD~152384 in the $J$ and
$H$ band, with a similar instrument configuration and at a comparable
airmass (1.78) as the science frames, in order to calibrate the
telluric and instrumental absorption profile.

The OSIRIS data reduction pipeline (DRP) was used to process and
extract the raw, two-dimensional (2D) spectra and place them back into
their spatial positions, producing a data cube of two spatial
dimensions $(x, y)$ and one wavelength dimension
($\lambda$)\citep{krabbe_etal04}.  As explained in
\citet{krabbe_etal04}, the DRP consists of a main processing routine
that calls data reduction modules in a sequential order to fully
reduce the data from the raw frames into a final data product.  The 2D
raw data were first processed by performing a pair-wise sky
subtraction, a removal of the spectral crosstalk associated with
extremely bright spectra on one row of the detector, an adjustment of
the 32 individual channels on the detector to remove any systematic
bias, a rejection of electronic glitches that occur during the readout
of the detector, and a cleaning of pixels overexposed by cosmic rays.
The above procedures eliminate instrumental artifacts prior to the
spectral extraction.  The most unique step within the OSIRIS pipeline
is the extraction of the spectra from the 2D raw frames.  This process
requires that the PSF of every lenslet position as a function of
wavelength has been mapped using a white light calibration lamp.
These PSFs appear to be stable over many months and the calibration is
performed infrequently either by the instrument team or Keck staff.
The extraction routine uses the spectral PSFs to iteratively assign
flux from a particular pixel into its corresponding lenslet spectrum.
Once the spectra are assigned to the appropriate lenslet, each
individual spectrum is resampled onto a regular wavelength grid using
linear interpolation and a global wavelength solution determined from
19 spectral arc lines.  Finally, the extracted spectra are inserted
into their respective spatial locations in a data cube.

\begin{figure}
\epsscale{0.9} \plotone{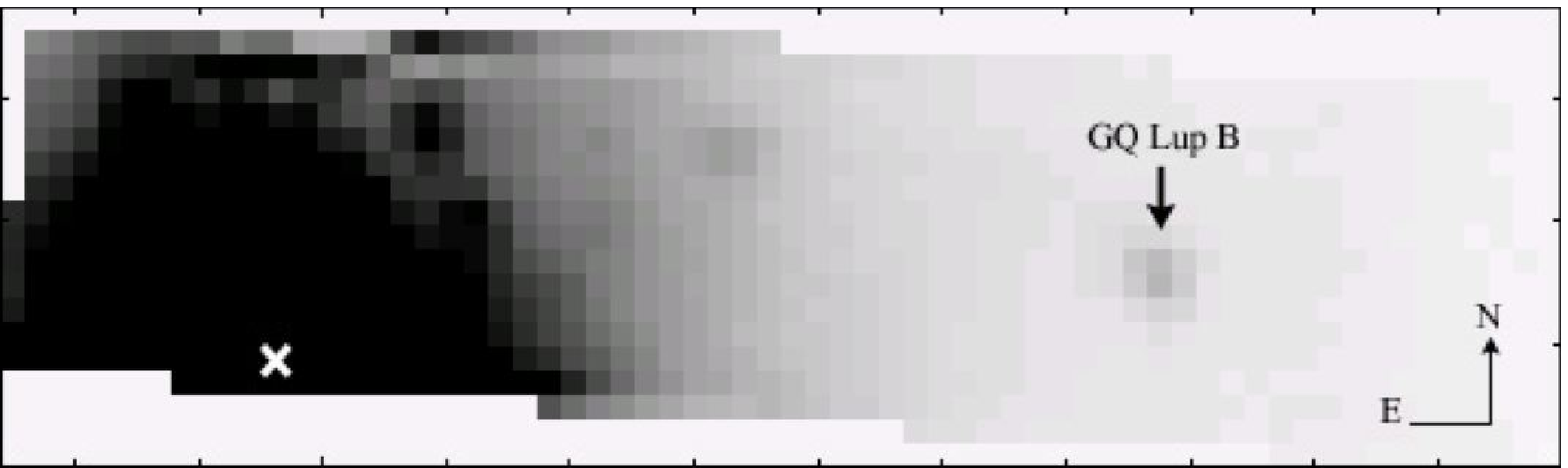} \figcaption{The median
collapsed $H$-band data cube of the GQ Lup system obtained on 2005
June 26.  The '$\times$' marks the position of GQ Lup A, and
0$\farcs$1 tick marks border the image.  GQ Lup B appears 7.5$\pm$0.5
mag fainter 0$\farcs$73$\pm$0$\farcs$01 away with a PA of
276.2$\pm$0.3$\degr$.
\label{fig_image}}
\end{figure}

A visual inspection of the $J$- and $H$-band data cubes confirmed the
companion GQ~Lup~B in the position identified by
\citet{neuhauser_etal05}.  The reduced $H$-band image of the GQ~Lup
system is displayed in Figure~\ref{fig_image}.  The fully reduced data
cube exhibits the effects of differential atmospheric dispersion, as
demonstrated by the spatial motions of the stellar location as a
function of wavelength.  Therefore, the profile of the differential
dispersion was calculated by measuring the centroid in each of the
spectral channels and fitting these data with a second order
polynomial.  The telluric spectra were obtained by measuring their
centroids at the front of the data cube (shortest observing $\lambda$)
and extracting from this center along the dispersion direction using a
2~spaxel (0$\farcs$040) radius circular aperture.  We correct for the
intrinsic features in the A0V spectra by fitting Lorentzian profiles
to each of the hydrogen Paschen and Brackett lines and subtracting
these fits from the original telluric spectrum.  The telluric and
instrument absorption profiles were removed from the final data cube
by dividing each spatial location by the normalized telluric spectrum
and multiplying by a normalized blackbody curve of $T_{\rm
eff}$=10000~K (corresponding to spectral type A0V).

The spectra of GQ~Lup~A and B were extracted from the telluric
corrected data cubes using the same technique that was applied to the
telluric standard.  However, at a separation of just 0$\farcs$7 the
spectrum of GQ~Lup~B was significantly contaminated by the bright halo
of its host star.  We approximated the background at the companion
location by extracting spectra of the halo of GQ~Lup~A at 1~spaxel
($0\farcs020$) intervals along a radial line connecting the host and
companion.  We then fit a third-order polynomial to each spectral bin
as a function of radial separation, and used the interpolated halo
spectrum at the distance of the companion as an estimate of the halo
contamination in the spectrum of GQ~Lup~B.  The halo contamination was
subsequently subtracted from the extracted spectrum of GQ Lup B.  The reduced $J$-
and $H$-band spectra of GQ~Lup~B are shown in
Figures~\ref{fig_jspec_comp} and \ref{fig_hspec_comp}.  

\section{RESULTS \label{sec_results}}

\subsection{The Spectrum of GQ~Lup~B \label{sec_spectrum}}

\subsubsection{$J$ Band \label{sec_spectrum_j}}

\begin{figure}
\epsscale{1.0} \plottwo{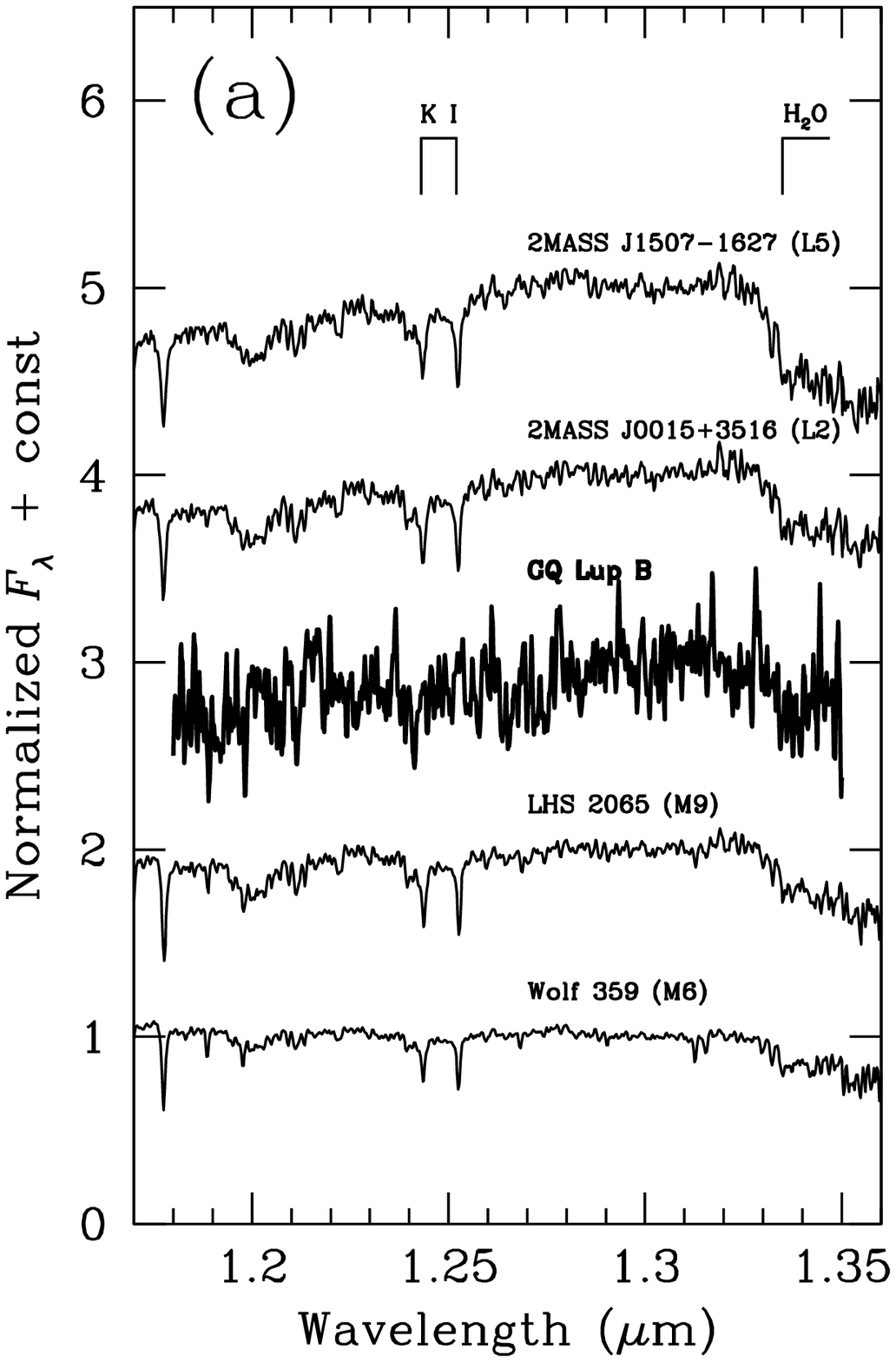}{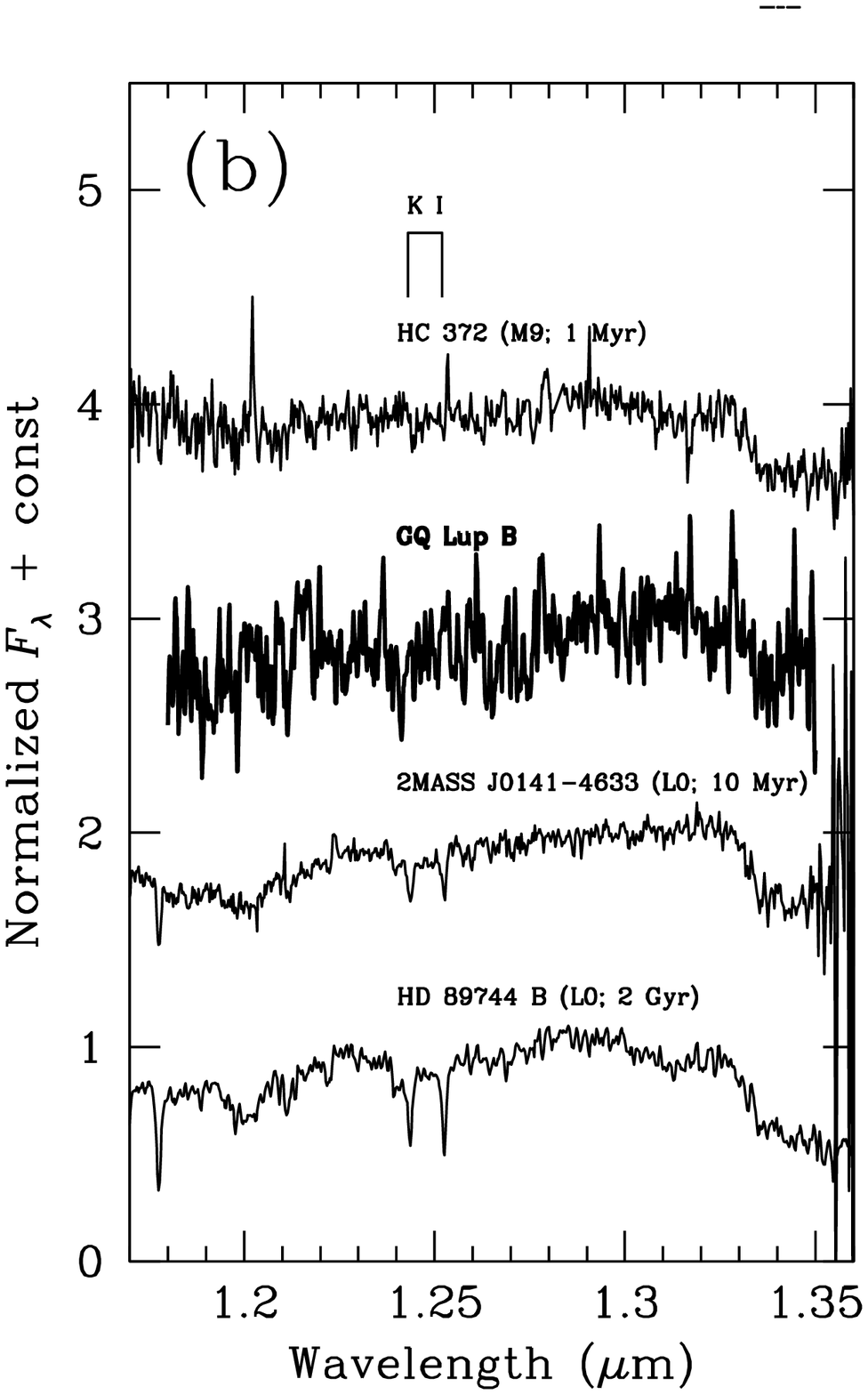} \figcaption{Comparison of
our 1.20--1.35~$\micron$ spectrum of GQ~Lup~B to the spectra of dwarfs
of a range of spectral types (a) and ages (b). The comparison spectra
in panel (a) are of field dwarfs from the NIRSPEC brown dwarf
spectroscopic survey of \citet{mclean_etal03}.  The comparison spectra
in panel (b) are from \citet[][HD~89744B]{mclean_etal03},
\citet[][2MASS~J01415823--4633574]{kirkpatrick_etal06}, and
\citet[][HC~372]{slesnick_etal04}.  The strength of the H$_2$O
absorption longward of 1.33~$\micron$ indicates a spectral type of
M6--L0 (\S\ref{sec_spectrum}).  Unlike older field dwarfs, but
similarly to the 1~Myr-old Orion Nebular Cluster dwarf HC~372,
GQ~Lup~B does not exhibit any \ion{K}{1} absorption at 1.243 and
1.252~$\micron$---an indication of low surface gravity and youth.  All
spectra are normalized to unity at 1.30~$\micron$.
\label{fig_jspec_comp}}
\end{figure}

\begin{figure}
\epsscale{0.75}
\plotone{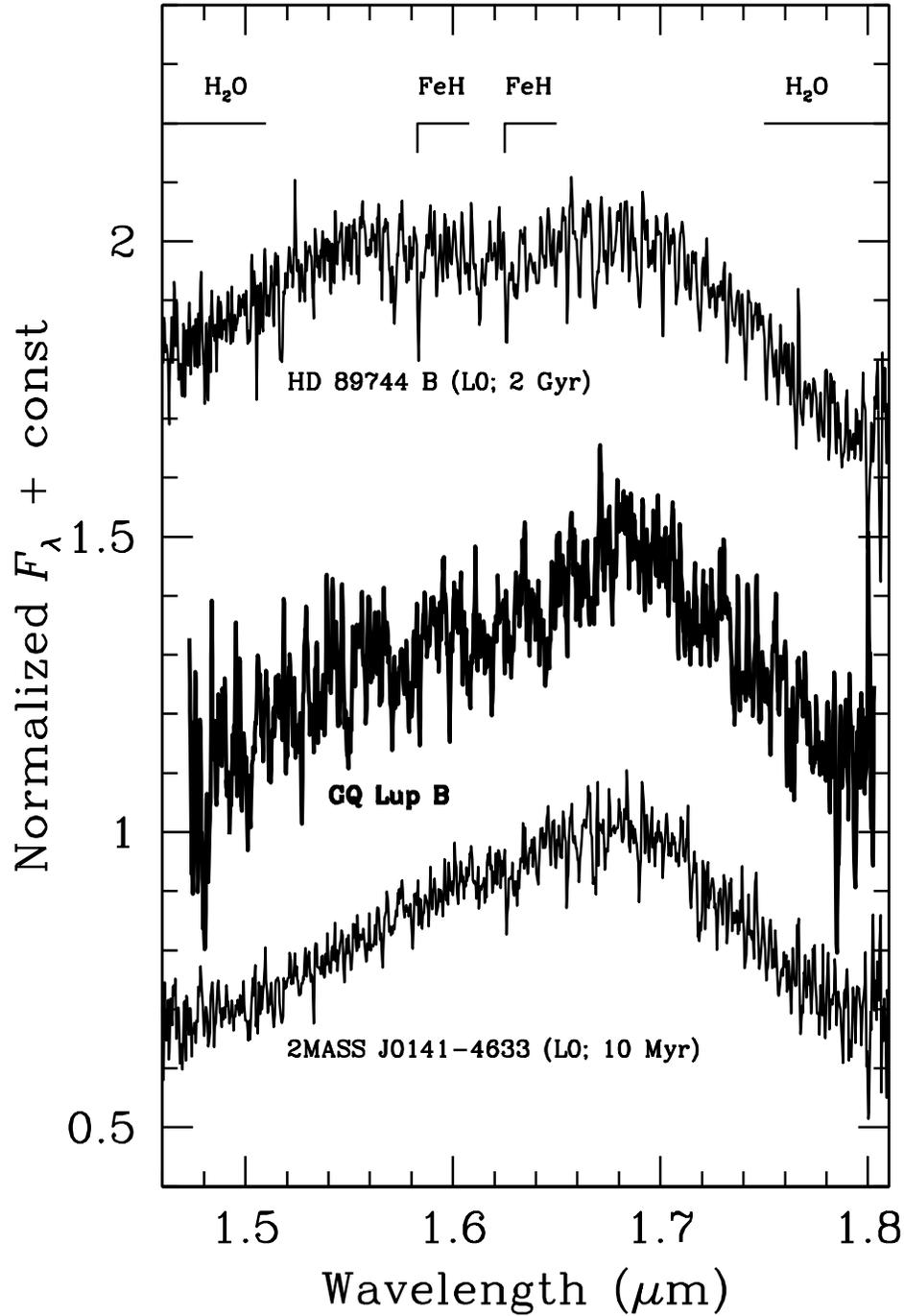}
\figcaption{The $H$-band spectrum of GQ~Lup~B compared to the spectra of L0 dwarfs of different ages.  The peaked continuum shape of the spectrum of GQ~Lup~B strongly resembles that of the $\sim$10~Myr old 2MASS~J01415823--4633574, indicating similarity in low surface gravity and spectral type.  All spectra are normalized to unity at 1.68~$\micron$.
\label{fig_hspec_comp}}
\end{figure}

We compare our $J$-band spectrum of GQ~Lup~B to the spectra of field
M6--L5 dwarfs from the NIRSPEC brown dwarf spectroscopic survey of
\citet{mclean_etal03} in Figure~\ref{fig_jspec_comp}a.  Our
$R$$\approx$3700 spectrum from OSIRIS has been smoothed to the
$R$$\sim$2000 resolution of the comparison NIRSPEC spectra.  The
spectrum of GQ~Lup~B exhibits a depression due to water absorption
longward of 1.33~$\micron$ consistent with the spectra of other M6--L5
dwarfs in the field.  However, unlike the several Gyr-old field
dwarfs, GQ~Lup~B does not show \ion{K}{1} absorption at 1.243 and
1.254~$\micron$---an indication that GQ~Lup~B has low surface gravity.
A comparison with $R$$\approx$2000 $J$-band spectra of M9--L0 dwarfs
of various ages exemplifies this point (Fig.~\ref{fig_jspec_comp}b).
With its lack of \ion{K}{1} absorption at $J$ band (3$\sigma$ upper
limit of 1.2~\AA\ on the equivalent width at 1.244 and
1.253~$\micron$), GQ~Lup~B closely resembles the $\sim$1~Myr M9 dwarf
HC~372 in the Orion Nebular Cluster \citep{slesnick_etal04}, but not
the older L0 dwarfs 2MASS~J01415823--4633574
($\sim$10~Myr\footnote{\citet{kirkpatrick_etal06} report an age of
1--50~Myr for 2MASS~J01415823--4633574, but they compare their derived
values of T$_{eff}$ and log ($g$) to the theoretical models of
\citet{baraffe_etal01} and find their best guess age estimate is
5--10~Myr.}; \citealt{kirkpatrick_etal06}; EW(\ion{K}{1}) =
3.5$\pm$0.5~\AA\ at each of the two central wavelengths) and HD~89744B
\citep[$\sim$2~Gyr;][]{wilson_etal01}.  The low surface gravity of
GQ~Lup~B indicates that it is a very young brown dwarf, as we later
confirm from an age analysis of its pre-main sequence primary
(\S\ref{sec_age}).

A visual comparison of the $J$-band continuum shapes of GQ~Lup~B and
the M6--L5 dwarfs in Figure~\ref{fig_jspec_comp}a indicates that the
spectral type of GQ~Lup~B is intermediate between those of mid-M and
early-L dwarfs.  We test this result by calculating the $J$-band
H$_2$O indices of \citet{mclean_etal03} and \citet{slesnick_etal04},
which measure the onset of water absorption at 1.34~$\micron$.  The
strength of water absorption has been reported to be gravity-sensitive
in the $H$-band, as displayed in the peaked shapes of $H$-band
continua of young ultra-cool dwarfs \citep{lucas_etal01,
luhman_etal04}.  However, \citeauthor{slesnick_etal04} find that at
1.34~$\micron$ water absorption is unaffected by surface gravity in
ultracool dwarfs, and is thus an adequate indicator of effective
temperature.  We verify this claim by applying the $J$-band H$_2$O
index of \citeauthor{slesnick_etal04} to the low-gravity L dwarf
2MASS~J01415823--4633574 \citep{kirkpatrick_etal06}, and obtain a
formal spectral type estimate of M9.5$\pm$1.1 for that object, fully
consistent with the $\sim$L0 classification of
\citet{kirkpatrick_etal06}.  Thus, using the $J$-band H$_2$O indices of
\citeauthor{mclean_etal03} and \citeauthor{slesnick_etal04}, we infer
a spectral type of M6--L0 for GQ Lup B.  The range in spectral type is
determined by observing the variations in the inferred spectral type
as a result of small shifts (up to 50~\AA) in the centers of the
spectral windows used in the index definitions.  This method allows a
better sampling of the noise in our spectrum.

\subsubsection{$H$ Band}

The $H$-band spectrum of GQ~Lup~B is shown in
Figure~\ref{fig_hspec_comp}, where it is compared to the spectra of
the $\sim$10~Myr old 2MASS~J01415823--4633574 \citep{kirkpatrick_etal06},
and the $\sim$2~Gyr old HD~89744B \citep{wilson_etal01}.  The sharply
peaked, triangular continuum of the spectrum of GQ~Lup~B strongly
resembles that of the young L0 object, while both differ from the
plateau-shaped continuum of the older L0 dwarf.  Similarly shaped
$H$-band continua have been reported in 1--10~Myr-old late-M/early-L
dwarfs before, and are now known to be indicators of low surface
gravity and youth \citep{lucas_etal01, luhman_etal04}.  The strong
gravity-dependence of the $H$-band continuum shape and the lack of
other obvious spectroscopic features \citep[though FeH absorption at
1.625~$\micron$ can be seen in the higher signal-to-noise spectrum of
2MASS~J01415823--4633574;][]{kirkpatrick_etal06} prevent us from using
the $H$-band spectrum of GQ~Lup~B for accurate spectroscopic
classification.  We only note that the overall similarity between the
spectra of GQ~Lup~B and 2MASS~J01415823--4633574 indicate proximity in
spectral type---in agreement with our $J$-band spectroscopic
classification (\S\ref{sec_spectrum_j}).

\subsection{Astrometry and Photometry of GQ Lup A and B}

In addition to the spectral information obtained from the OSIRIS data
cube, typical imaging measurements can be performed on the
three-dimensional data cube.  Relative astrometry of the binary was
calculated through measurements of the component centers in collapsed
narrowband (100 spectral channel bins) images for each of the two
$J$-band frames that contained both GQ Lup A and B.  We fit 2D
gaussian profiles to the clearly resolved cores of the PSFs of each
component in order to attain their respective location in the frame.
We measure the separation between the components to be
0$\farcs$73$\pm$0$\farcs$01 with a position angle of
276.2$\degr$$\pm$0.3$\degr$.  Our astrometric observations are fully
consistent with the values reported in \citet{neuhauser_etal05}, and
we agree that GQ Lup A and B comprise a common proper motion system.

Magnitudes for the system were obtained by comparing the relative
spectral intensities of the 2 spaxel radius extracted spectra.  It was
necessary to derive a flux for GQ Lup A instead of using the published
2MASS value because this source is variable at many wavelengths
(\S\ref{sec_age}).  We calibrate our photometry with the telluric
standard; however, since the telluric standard was intrinsically
brighter than GQ Lup A, the AO performance on the telluric standard was
better and therefore the core of the PSF was narrower.  Consequently,
we could not directly compare the total flux of the 2 spaxel radius
extracted spectrum of the telluric standard with that of GQ Lup A.  We
derive a radial profile curve of growth for both GQ Lup A and the
telluric standard, and we approximate the total enclosed flux by
assuming circular symmetry.  The magnitudes were estimated from 30
spaxel radius apertures, which alleviates the effect of different PSF
shapes.  In the $J$-band, we use the 2MASS magnitude for the telluric
standard (HD~152384) to calibrate the system photometry.  Our derived
$J$-band magnitude for GQ Lup A is $J$=8.69$\pm$0.04 mag, which is
consistent with the 2MASS quoted value of 8.605$\pm$0.021, meaning
that the system is in the same photometric state.  However, in the
$H$-band, we find that our derived magnitude for GQ Lup A is
$H$=8.0$\pm$0.2 mag, a value significantly larger than the 2MASS value
of 7.702$\pm$0.033 mag.  This inconsistency is probably due to errors
in fitting the radial profile for GQ Lup A since the $H$-band center
of GQ Lup A is on a spaxel at the edge of the field of view (see
Figure~\ref{fig_image}), instead of some intrinsic color variability
of GQ Lup A.  For this reason, we calculate the GQ Lup B $H$-band
magnitude using the 2MASS magnitude quoted for GQ Lup A, but we
propagate an appropriate error term to account for this discrepancy.
We find that GQ Lup B has $J$=14.90$\pm$0.11 mag
($\Delta$$J$=6.21$\pm$0.12 mag) and $H$=15.2$\pm$0.5 mag
($\Delta$$H$=7.5$\pm$0.5 mag).  As mentioned above, these near-IR
magnitudes are unique to the OSIRIS instrument where each filter
corresponds to the different orders of the diffraction grating.
Following \citet{ste04}, we estimate that for an object of spectral
type M6--L0, an error of $\sim$0.1 magnitudes is necessary to
translate to the standard near-IR standard filters in the $J$ or
$H$-band, and this error has been added to our uncertainty.

\section{DISCUSSION \label{sec_analysis}}

In order to determine the luminosity and model-dependent mass of
GQ~Lup~B, we need accurate estimates of its distance from the Sun and
its intrinsic age.  We will assume throughout this discussion that
GQ~Lup~B is gravitationally bound to GQ~Lup~A, which implies that
it resides at an identical heliocentric distance as the primary.  We also
assume that physical association in the binary implies coeval formation.

\subsection{Comparison with the Spectral Type Inferred by \citet{neuhauser_etal05} \label{sec_neu_comparison}}

The final spectral type of GQ~Lup~B is consistent with, albeit
somewhat earlier than, the M9--L4 determination from $K$-band
spectroscopy in \citet{neuhauser_etal05} and \citet{guenther_etal05}.
The estimates in these two papers are primarily based on the continuum
slope around 2.0~$\micron$, where spectra of ultra-cool dwarfs exhibit
H$_2$O absorption.  The effect of surface gravity on this absorption
band has not been empirically determined, and it is possible that
enhanced water absorption in low surface gravity photospheres may be
depressing the continuum near 2~$\micron$, thus making objects appear
cooler.  Such an interpretation draws an analogy with the perceived
role of water absorption in creating the peaked $H$-band continua of
young ultra-cool dwarfs \citep{luhman_etal04}.  However, the lack of
sensitivity to surface gravity of the 1.34~$\micron$ water absorption
band \citep{slesnick_etal04} is perplexing in this context.

A more self-consistent picture of the shapes of the near-IR spectra of
young ultra-cool dwarf near-IR spectral shape has been recently
offered by \citet{kirkpatrick_etal06}.  These authors argue that,
rather than due to enhanced water absorption, the triangular-shaped
continua at $H$-band are caused by a reduction in H$_2$ collision
induced absorption (CIA) at low surface gravity
\citep{borysow_etal97}.  \citet{borysow_etal97} show that at the
temperatures of late-M and early-L dwarfs ($\sim$2300--2500~K) H$_2$
CIA peaks in strength near 2.5~$\micron$, and weakens toward shorter
wavelengths, or with decreasing surface gravity.
\citet{kirkpatrick_etal06} find that the theoretical picture of
\citet{borysow_etal97} correctly predicts two main features of young
late-M/early-L dwarfs: redder $J-K_S$ colors and peaked $H$-band
continua.  Indeed, employing the $K_{s}$ magnitude from
\citet{neuhauser_etal05} with the photometry presented herein, we find
a $J$-$K_{s}$ color of 1.8$\pm$0.1 mag, significantly redder than the
$J$-$K$$_{s}$=1.0--1.2 for typical field M6-L0 dwarfs \citep{leg02}.
Given the decreasing strength of H$_2$ CIA toward shorter wavelengths,
this interpretation also explains the diminished sensitivity to
surface gravity in the depth of the 1.34~$\micron$ water absorption
band in ultra-cool dwarfs.  The effect is strongest at $K$-band, where
the decreasing strength of H$_2$ CIA with decreasing surface gravity
makes the continuum redder.  In particular, $K$-band spectroscopic
classification of young ultra-cool dwarfs based on the water-band
continuum slope near 2~$\micron$ will produce later spectral types
than other classification schemes (e.g., spectral types based on the
strength of the 1.34~$\micron$ water band).  Hence, this scenario
offers an explanation of the slightly later spectral type obtained for
GQ~Lup~B by \citet{neuhauser_etal05} and \citet{guenther_etal05}.

A comparison of the absolute magnitude of GQ~Lup~B with other
similiarly young brown dwarfs suggests that GQ~Lup~B is, in fact, of
an earlier spectral type than $\sim$L0.  At an $M_{K_S}=7.2\pm0.3$~mag
(adopting a distance of 150+/-20 pc to GQ Lup; \S\ref {sec_dist}),
GQ~Lup~B is 1--2~mag brighter than the 1--10~Myr $\sim$M9.5 dwarfs
OTS~44 ($M_{K_S}=8.48$~mag) and Cha~110913--773444 ($M_{K_S}=9.6$~mag)
in Chamaeleon \citep{luhman_etal04,luhman_etal05b}, and the 1--5~Myr
old M9/L0 binary Oph~162225--240515~A/B
\citep[$M_{K_S}=8.19/8.75$~mag;][]{jayawardhana_ivanov06} in
Ophiuchus.  In addition, GQ~Lup~B is also of approximately the same
bolometric luminosity ($\log L/\Lsun=-2.2$; \S\ref{sec_mass}) and age
as the 3--5~Myr M7--M7.5 dwarfs USco~128 and USco~130 \citep[$\log
L/\Lsun=-2.4$;][]{mohanty_etal04b} in Upper Scorpius.  Therefore, our
mean spectral type estimate of M8 for GQ~Lup~B is in agreement with
expectations when compared to the spectral types of similar objects
from the literature.

A different explanation for the discrepancy between the spectral types
of GQ~Lup~B inferred in the present work and in
\citet{neuhauser_etal05} may be suggested as a result of previously
noted difficulties in reproducing the correct continuum slopes of
objects observed with AO long-slit spectroscopy \citep{goto_etal03}.
The width of spectroscopic slits used with AO is often 1--2 times the
FWHM of the PSF ($\sim$60~mas at 2.2~$\micron$ on 8--10~meter class
telescopes), and of order of the accuracy ($\approx$20~mas) with which
an object can be positioned and maintained on the slit during dithers.
Because the PSF possesses a strong radial chromatic gradient, any
misalignment of the target on the slit can lead to an artificial
change in the measured continuum.  Furthermore, in high-contrast
observations of binary systems (i.e., when one object is much fainter
than the other) it is desirable to have the slit oriented along the
binary axis to allow accurate determination of the contamination of
the secondary spectrum by the halo of the primary.  As a result,
spectroscopy of binaries rarely benefits from having the slit oriented
along the parallactic angle, and can suffer additional slit losses due
to differential atmospheric refraction (DAR), especially at high
($\gtrsim$1.5) airmasses and short ($\lesssim$1.5$\micron$)
wavelengths.

Given that \citeauthor{neuhauser_etal05} observed GQ~Lup at $K$ band
and at low ($<$1.1) airmass, the above described slit losses due to
DAR are negligible.  However, the effect of potential misalignment of
GQ~Lup~B and the slit could be significant.  \citet{goto_etal03} find
that a misalignment equal to half the slit width changes the measured
continuum slope by 7--8\% per $\micron$ at $H$ band.  The effect is
likely less pronounced at $K$ band, and the width of the slit used by
\citeauthor{neuhauser_etal05} (172~mas) is larger than that of the one
used by \citeauthor{goto_etal03} (100~mas) in their experiment.
Hence, maintaining adequate alignment on the slit should have been
easier to achieve in the case of the $K$-band spectroscopic
observations of GQ~Lup~B by \citeauthor{neuhauser_etal05}.  Therefore,
we conclude it is improbable that slit losses have significantly
altered their $K$-band continuum shape, and that the likely reason for
the discrepancy between the spectral types of GQ~Lup~B inferred by
\citeauthor{neuhauser_etal05} and in the present work remains H$_2$
CIA.

\subsection{The Heliocentric Distance to GQ Lup \label{sec_dist}}

GQ~Lup~A is a T~Tauri star located in the Lupus~1 star-forming region
\citep{schwartz77}.  With a visual magnitude of $V\approx12$~mag
\citep{covino_etal92, gregoriohetem_etal92}, it is too faint to have a
trigonometric parallax measurement from {\sl Hipparcos}.  However, its
heliocentric distance can be inferred from the distance to its parent
molecular cloud; as determined from interstellar reddening,
polarization, or \ion{Na}{1} absorption in the spectra of objects
along the same line of sight, or from the mean distance to brighter
early-type members of the cloud.  \citet{neuhauser_etal05} and
\citet{guenther_etal05} summarize the available distance estimates for
GQ~Lup from the literature, and converge on a value of 140$\pm$50~pc.
The adopted distance correctly represents the entire range of
published distances to the overall Lupus star-forming region
\citep[100--190~pc;][]{hughes_etal93, wichmann_etal98, knude_hog98,
teixeira_etal00}.  A more precise estimate can be obtained from a
joint comparative analysis of the techniques employed in the various
studies, and by discriminating among the four sub-groups
\citep{schwartz77} of the Lupus molecular cloud complex.  Such a
comprehensive analysis is presented in \citet{franco02}, who concludes
that the mean distance to the entire molecular cloud complex, and to
Lupus~1 in particular, is $\approx$150~pc.  According to
\citet{franco02}, the near edge of the Lupus~1 cloud is at least
130--140~pc away, based on studies of interstellar polarization
\citep{rizzo_etal98} and \ion{Na}{1} absorption \citep{crawford00}
toward Lupus~1.  Hence, we conservatively adopt a 20~pc uncertainty in
the inferred 150~pc distance to Lupus~1.

\subsection{The Age of GQ~Lup \label{sec_age}}

From a comparison to stellar evolutionary models,
\citet{neuhauser_etal05} find that the age of GQ~Lup~A is between
0.1~Myr and 2~Myr.  While this is less than the median age (3~Myr) of
the Lupus star-forming region \citep{hughes_etal94}, this region
contains both classical and weak-line T~Tauri populations that may
have different ages.  The Lupus~1 and 2 sub-regions, in particular,
have the highest concentration of classical T Tauri stars and are
considered to be the youngest \citep[0.1--1~Myr;][]{hughes_etal94}.
The location of GQ~Lup in Lupus~1 indicates that indeed an age younger
than 3~Myr may be warranted.  However, we note that GQ~Lup has
alternately been classified both as a weak-line T~Tauri star
\citep[H$\alpha$ equivalent width of 2.8~\AA;][]{herbig_bell88} and as
a classical T~Tauri star (H$\alpha$ equivalent width of 38.6~\AA;
\citealt{appenzeller_etal83}; \citealt{hughes_etal94}).  The variation
in the strength of its H$\alpha$ emission precludes a conclusive
association with either the classical or weak-line populations in
Lupus.  Therefore, we proceed to derive an independent estimate of the
age of GQ~Lup.

We use $R$- and Cousins $I$-band photometry from the literature to
place GQ~Lup on a $M_I$ vs.\ $R-I$ color-magnitude diagram and compare
its position to the predictions of theoretical models for
pre-main-sequence stars.  The $R$ and $I$ bands are not strongly
affected by excess UV and IR emission, and are thus suitable proxies
for the bolometric luminosities and effective temperatures of
pre-main-sequence stars.  $R$- and $I$-band photometry of
pre-main-sequence stars is also often obtained
simultaneously\footnote{$I$- and $J$-band photometry are another
suitable filter pair for estimating bolometric luminosities and
effective temperatures, but these filter complements require two
different detectors.}, and thus allows self-consistent measurements of
the color of variable stars.  This is a particularly important
consideration in the case of GQ~Lup, which is strongly variable
\citep[up to 2~mag at $V$;][]{covino_etal92}.  GQ~Lup is further known
to have non-negligible visual extinction, where reports vary in the
literature---$A_V=0.4\pm0.2$~mag \citep{batalha_etal01},
$A_V=0.95$~mag \citep{hughes_etal94}, or $A_V=1.6$~mag
\citep{bertout_etal82}.  Barring systematic differences in the
approaches used to determine the visual extinction in the three cases,
the variability in visual extinction may be linked to the photometric
variability of GQ~Lup, pointing to a probable circumstellar origin for
the extinction.  Such a conclusion is supported by simultaneous
$UBVRI$ monitoring observations of GQ~Lup, from which
\citet{covino_etal92} find that the changes in the optical colors
approximately follow a standard interstellar extinction law.  We adopt
the mean of the above extinction values and their standard deviation,
i.e., $A_V=1.0\pm0.6$~mag, as representative of the extinction toward
GQ~Lup at any given time, and use the interstellar extinction law of
\citet{cardelli_etal89} to convert $A_V$ to $R$- and $I$-band
extinction.  GQ~Lup is also known to exhibit variable veiling in its
optical spectrum \citep{batalha_etal01}, with the amount of excess
emission ($F_{\rm ex}$) being 0.5--1.5 times the photospheric level
($F_{\rm phot}$) at $V$-band, and 1.0--4.5 times at $B$-band.  Veiling
of the optical continuum in T Tauri stars occurs as a result of excess
emission at the location where the accretion column collides with the
stellar surface in a radiative shock.  Adopting a simple blackbody
model for the veiling with the derived effective temperatures in
\citet{batalha_etal01}, we find that the veiling of GQ~Lup makes, on
average, the $I$-band photometry 0.35~mag brighter (amplitude of
variation:  $\pm$0.24~mag) and the $R-I$ color 0.15~mag bluer (amplitude of
variation:  $\pm$0.07~mag).

\begin{figure}
\epsscale{0.5}
\plotone{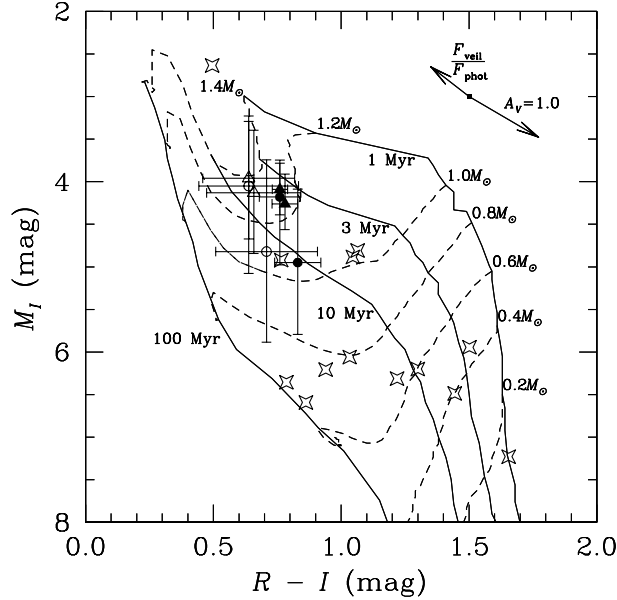}
\figcaption{A color-magnitude diagram of GQ Lup (data points with errorbars) and
other Lupus~1 members (open star symbols) with evolutionary tracks from
\citet{baraffe_etal98}.  The solid symbols represent the observed data
from four photometric data sets before de-reddening and de-veiling.  The
solid circles show the mean magnitudes and colors of long-period
($>$10~days) photometric campaigns from \citet{covino_etal92} and
William Herbst (unpublished; available at {\sl
ftp://ftp.astro.wesleyan.edu/ttauri}).  The solid triangles show
single-epoch measurements from \citet{gregoriohetem_etal92} and
\citet{hughes_etal94}.  The errorbars on the solid symbols denote the
quadrature sum of the full range of observed photometric variation and
the 1$\sigma$ uncertainty (0.32~mag) in the distance modulus of GQ~Lup.
The open circles and triangles indicate the de-reddened and de-veiled
data for GQ~Lup, using the mean reddening and veiling estimates from
\S\ref{sec_age} (shown as vectors in the upper right corner of the
Figure).  The errorbars on the open symbols include the full amplitudes
of the inferred reddening and veiling.  The star-like symbols represent
other de-reddened and de-veiled Lupus~1 members from
\citet{hughes_etal94}.  For these, we have adopted visual extinctions
from \citet{hughes_etal94}, and the veiling vector for GQ~Lup.
\label{fig_cmd}}
\end{figure}

Figure~\ref{fig_cmd} shows the photometric measurements of GQ~Lup from
each of four separate optical data sets.  Solid circles and solid
triangles show the original data, without de-reddening and de-veiling,
whereas the open circles and open triangles correspond to the de-reddened
and de-veiled data.  The error bars on the photometry and colors
correspond to the quadrature sum of the amplitude of the observed
variability and the 1$\sigma$ uncertainty (0.32~mag) in the distance
modulus.  For the de-reddened and de-veiled data we also include the
full ranges of the inferred reddening and veiling.  Also overlaid are
isochrones (solid curves) and evolutionary tracks (dashed curves) from
\citet{baraffe_etal98} with the mixing length parameter $\alpha=1.0$.
In a comparative analysis of theoretical evolutionary tracks,
\citet{hillenbrand_white04} find that this set of models most
accurately predicts the dynamical masses of pre-main-sequence stars.

As is evident in Figure~\ref{fig_cmd}, the photometric data reveal
that the age of GQ~Lup is $\approx$3~Myr, although ages from 1~Myr to
$\sim$50~Myr are within the allowed $M_I$ and $R-I$ locus.  An age
$>$10~Myr can be excluded if we treat GQ Lup A as a classical T~Tauri
star which are not typically found in associations older than
$\sim$10~Myr \citep[the approximate age of the TW~Hya
association;][]{kastner_etal97, mamajek05}.  Indeed, we argue that
because circumstellar extinction probably contributes significantly to
the observed variability, we need to weigh the data in
Figure~\ref{fig_cmd} more heavily toward higher luminosities (i.e.,
younger ages), which likely represent the unextincted state.  We
therefore conclude that GQ~Lup is $\approx$3~Myr old, with a possible
age range of 1--10~Myr, i.e., marginally older than the 1$\pm$1~Myr
estimate of \citet{neuhauser_etal05}.

\subsubsection{A Need to Revise the Age of Lupus?}

Our age estimate for GQ~Lup from the evolutionary tracks of
\citet{baraffe_etal98} suggests an older age for the Lupus~1
star-forming region as a whole, than the currently quoted value of
0.1--1~Myr.  The latter age range was derived by \citet{hughes_etal94}
based on models from \citet{dantona_mazzitelli94}, which have now been
shown to systematically under-estimate the masses of pre-main-sequence
stars \citep{hillenbrand_white04}.  Indeed, at an inferred age of
0.1~Myr, GQ~Lup is ranked as one of the youngest stars in Lupus
according to \citeauthor{hughes_etal94} Our updated, older age for
GQ~Lup indicates that the entire region is probably $\approx$10 times
older than found by \citet{hughes_etal94}, i.e., 1--10~Myr.  The
color-magnitude diagram in Figure~\ref{fig_cmd} contains all fourteen
Lupus~1 members listed in \citet{hughes_etal94}.  For all stars we
have adopted the photometry and extinctions listed in Table~3 of
\citet{hughes_etal94} and the same veiling as for GQ~Lup.
Figure~\ref{fig_cmd} demonstrates that the majority of the Lupus~1
members are at least 3~Myr-old, with an age scatter of 1--100~Myr.
However, ages of $>$10~Myr would be highly unusual for any members in
this molecular cloud, which has a high incidence of classical T~Tauri
stars.  Barring major inaccuracies in the \citet{baraffe_etal98}
models, the old appearance of some of the stars could be explained by
local deviations from the adopted interstellar extinction law.
Indeed, because of dust grain growth in primordial circumstellar
disks, extinction and reddening due to circumstellar dust is not
always described adequately by the standard interstellar medium (ISM)
extinction law, which holds for grains 0.01--0.1~$\micron$ in size.
Instead, circumstellar extinction is often more neutral in color
(``gray''), especially in systems with close to edge-on viewing
geometries, where the light from the central source passes through a
non-negligible part of the disk \citep[e.g.,][]{throop_etal01}.
Hence, because \citet{hughes_etal94} determine extinctions based on
observed $R-I$ colors and a priori known spectral types, the inferred
ISM-like extinctions may under-estimate the actual ones.  As a result,
on a color-magnitude diagram the stars would appear fainter, but not
redder.  That is, the stars would appear older.  It is possible that
some of the de-reddened and de-veiled data points in
Figure~\ref{fig_cmd}, especially along the older isochrones,
under-estimate the actual unextincted stellar luminosities and ages.
An upper age limit of $\sim$10~Myr can be inferred by comparison to
the TW~Hya association, which is the oldest stellar association known
to harbor classical T~Tauri stars.  Considering the upper envelope of
the data, which presumably are least affected by gray extinction,
there is a strong reason to believe that the stellar population in the
Lupus~1 molecular cloud is $\geq$1~Myr old.

The above reasoning leads us to infer that the age of Lupus~1 is
1--10~Myr, or 10 times older on average than previously presumed.
Similar analysis leads to an analogous conclusion for the ages of
other young star-forming regions of ages comparable to that of Lupus 1
(e.g., Taurus).  In view of the continuous improvement in stellar
evolutionary models, a broad re-analysis of stellar ages in
star-forming regions may indeed be necessary in the future.  However,
given the remaining uncertainties in the theory at $\sim$1~Myr ages,
we can not claim a need for a significant re-evaluation of
pre-main-sequence ages, despite the relative success of the
\citet{baraffe_etal98} models of the Lyon group in reproducing
dynamical masses of pre-main-sequence stars.  We adopt these models to
estimate the age of GQ~Lup (and Lupus~1) in the present analysis to
ensure self-consistency with the evolutionary models \citep[][also
from the Lyon group]{chabrier_etal00} that we will use to estimate the
mass of GQ~Lup~B (\S\ref{sec_mass}).  In addition to being some of the
most widely used and successful (\S\ref{sec_mass}) sub-stellar
evolutionary models to date \citep[the other set coming from the
Arizona group;][]{burrows_etal97}, the sub-stellar evolutionary models
of \citet{chabrier_etal00} allow us to estimate the mass of GQ~Lup~B
based on the same theoretical framework used in estimating the age of
the primary.

\subsection{The Mass of GQ~Lup~B \label{sec_mass}}

We base our estimate of the mass of GQ~Lup~B on the the models of
\citet{burrows_etal97} and \citet{chabrier_etal00}.  We obtain its
bolometric luminosity using the derived $J$ magnitude, our distance
estimate (\S\ref{sec_dist}), and $K$-band bolometric corrections
for M6--L0 dwarfs from \citet{golimowski_etal04}.  Although based
solely on optical--near-IR data, the values of the bolometric
corrections have been largely confirmed by recent 5.5--38~$\micron$
{\sl Spitzer}/IRS spectra of ultra-cool dwarfs \citep{cushing_etal06}.
The bolometric corrections are translated from the $K$ to $J$ band
assuming the $J$-$K$ color for M6-L0 dwarfs is 1.1$\pm$0.1 mag
\citep{leg02}.  We note that the bolometric corrections in
\citet{golimowski_etal04} are compiled from data for $>$1~Gyr-old,
high-surface gravity field dwarfs, and may need to be corrected for
the expected $\approx$1~dex lower surface gravity of GQ~Lup~B.  The
sense and magnitude of this correction is unknown empirically, as the
body of data on young ultra-cool dwarfs is extremely limited.  We use
the models of \citet{chabrier_etal00} to infer that the correction to
BC$_J$ for a +1~dex change in surface gravity is $\approx$0.15~mag.
We adopt $\pm$0.10~mag as an error estimate for the surface gravity
correction to BC$_{J}$, i.e., of the same order as the precision of
empirical bolometric corrections.  We have not applied an extinction
correction to our $J$-band photometry.  At $J$-band the amount of
extinction is only a third of the visual extinction \citep[for an ISM
extinction law;][]{cardelli_etal89}, and given the adopted
$A_V=1.0\pm0.6$~mag toward GQ~Lup~A it would be $A_J=0.28\pm0.17$~mag,
i.e., the estimated luminosity of GQ~Lup~B would increase by
$\approx$30\%.  However, we choose not to apply this correction
because we concluded that the extinction toward GQ~Lup~A is probably
circumstellar in origin (\S\ref{sec_age}).  Given the relatively wide
separation ($\approx$110~AU) between the primary and the secondary,
and the lack of evidence (e.g., much higher circum-primary extinction)
for a high-optical-depth edge-on viewing geometry for the system, we
believe that near-IR light from the secondary is negligibly extincted
by dust in the circum-primary disk.

\begin{figure}
\epsscale{1.0} \plottwo{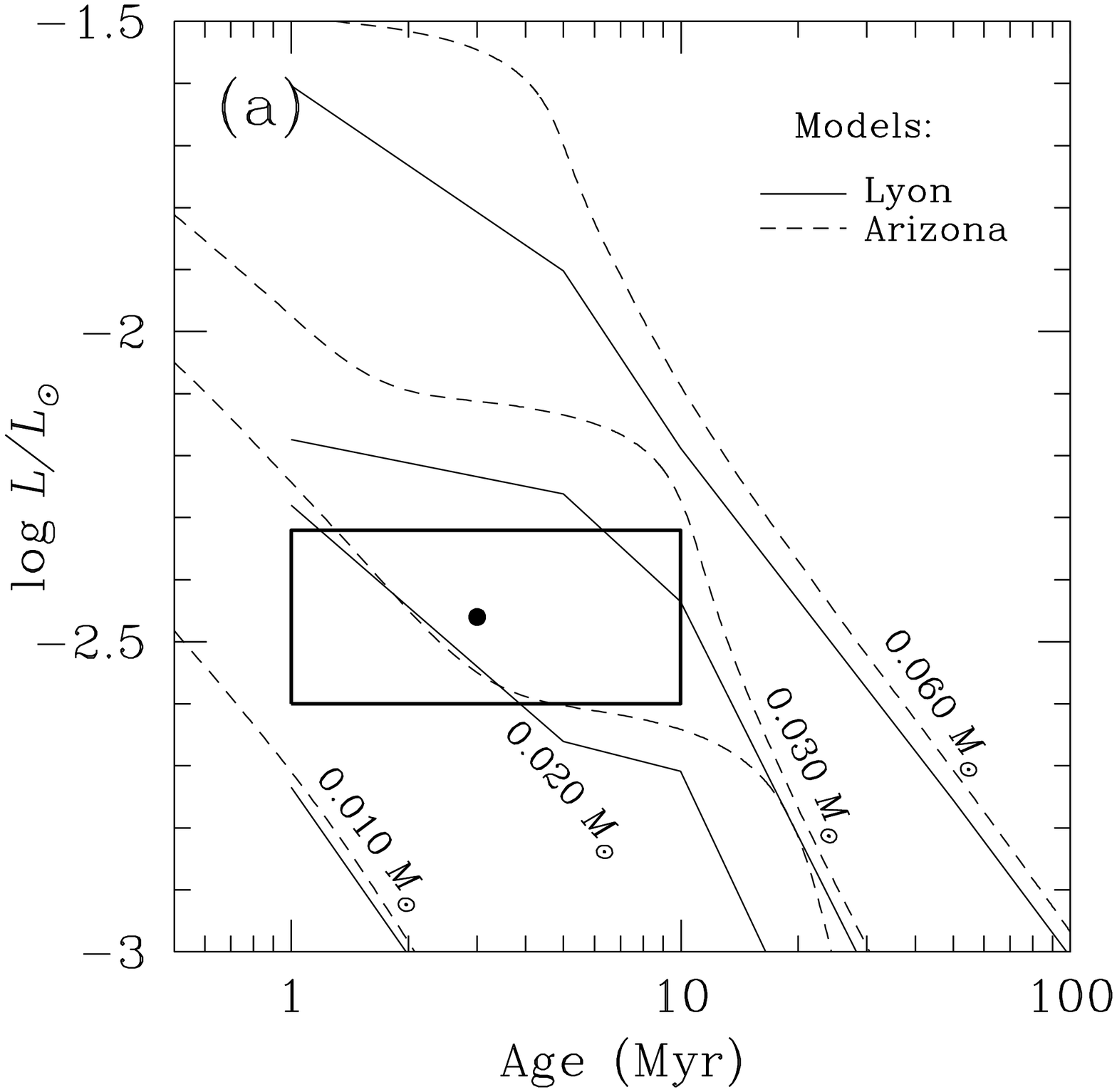}{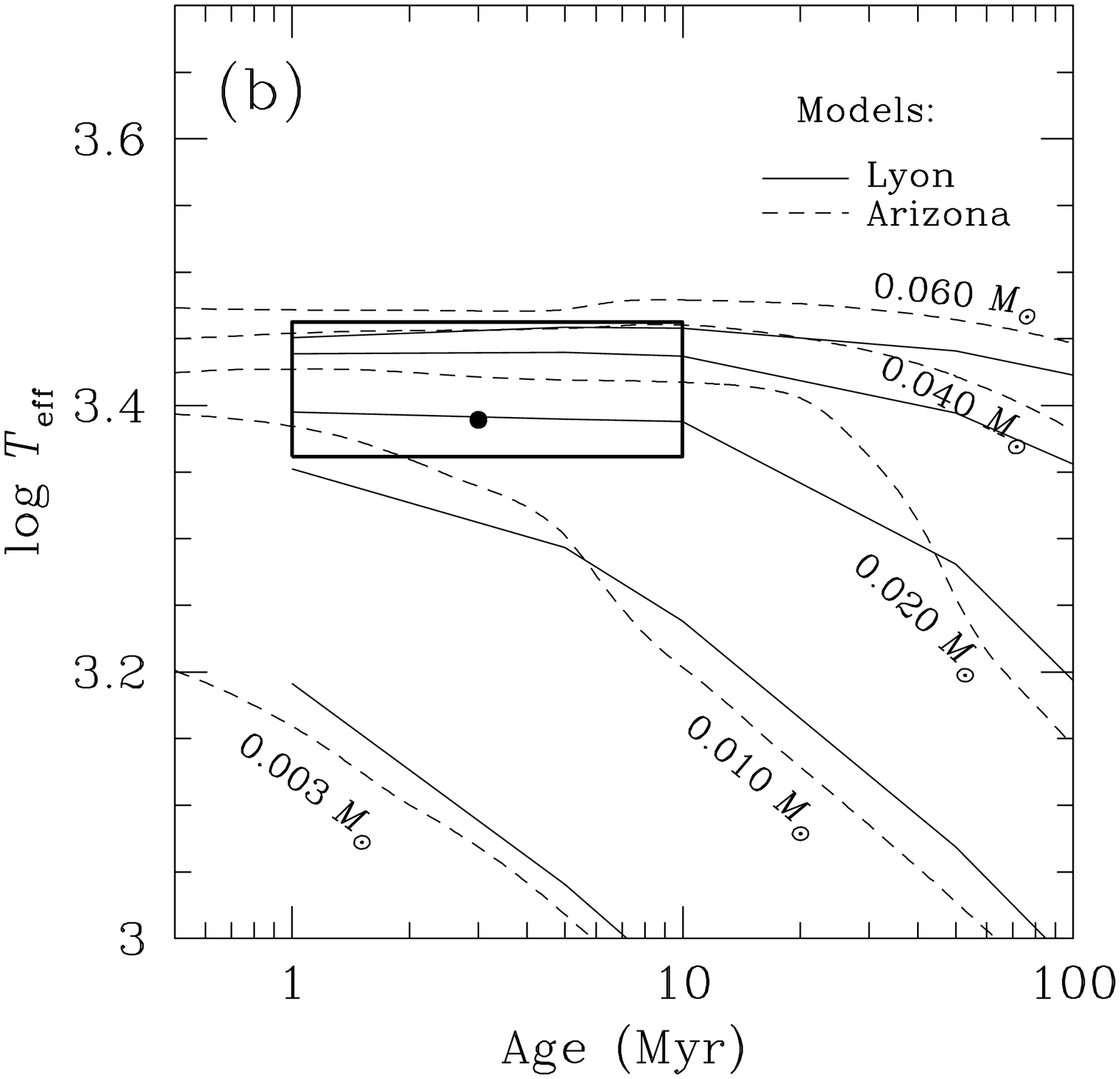}
\figcaption{Luminosity---age {\bf (a)} and effective temperature---age
{\bf (b)} evolution diagrams for GQ~Lup~B, with models from
\citet[][``Lyon'']{chabrier_etal00} and
\citet[][``Arizona'']{burrows_etal97} overlaid.  The solid dot in each
panel represents the mean estimate of the parameters of GQ~Lup~B,
while the thick rectangle delimits the allowed range of their
variation.  The predictions for the mass of GQ~Lup~B based on its
bolometric luminosity and effective temperature are consistent.  
\label{fig_mass}}
\end{figure}

The resulting estimate for the bolometric luminosity of GQ~Lup~B is
$\log L/L_\sun=-2.46\pm0.14$.  The effective temperature estimate is
2450~K with a range of 2300--2900~K, corresponding to the spectral
type M8$\pm$2, according to the T$_{eff}$/spectral type relation from
\citet{golimowski_etal04}.  Comparisons of these values with
sub-stellar evolutionary models from \citet{burrows_etal97} and
\citet{chabrier_etal00} are presented in Figure~\ref{fig_mass}.  We
estimate the mass of GQ~Lup~B at $\sim$0.012--0.040\Msun~based on its
bolometric luminosity, or $\sim$0.010--0.040\Msun~based on its
effective temperature.  Despite the use of the same photometry, our
mass estimate for GQ~Lup~B is a factor of $\sim$1.5--2 higher than
that of \citet[][based on the same models]{neuhauser_etal05}, largely
because of the older age that we estimate for the primary
(\S\ref{sec_age}).

We have not performed a detailed estimate of the mass of GQ~Lup~B
based on the models of \citet{wuchterl_tscharnuter03} as done in
\citet{neuhauser_etal05} because these models are not publicly
available.  Nevertheless, if we overlay the values for the effective
temperature and luminosity of GQ~Lup~B on an H-R diagram containing
tracks from \citeauthor{wuchterl_tscharnuter03} \citep[e.g., Figure~4
in][]{neuhauser_etal05}, we confirm the 1--2~\Mj\ mass estimate of
\citeauthor{neuhauser_etal05} This result is favored by
\citeauthor{neuhauser_etal05}, who argue that at the very young age of
GQ~Lup the models of \citeauthor{wuchterl_tscharnuter03} provide a
more realistic account of the collapse and formation of sub-stellar
objects.  \citeauthor{burrows_etal97} and
\citeauthor{chabrier_etal00} model sub-stellar evolution only
post-collapse, by assuming a pre-existing fully convective internal
structure which is adiabatic at all stages of evolution (i.e., a
``hot start'').  Such models are thus uncertain at ages up to a few
Myr \citep{baraffe_etal02}, and may be inadequate for the
$\sim$3~Myr-old GQ~Lup~B.  However, \citet{janson_etal06} argue that
the version of the \citeauthor{wuchterl_tscharnuter03} models used by
\citeauthor{neuhauser_etal05}, based on a core-accretion---gas-capture
scenario within a circumstellar disk, are also inappropriate at the
young age of GQ~Lup because they may not allow sufficient time for the
formation of the secondary.

The decision of which models to use in this scenario is therefore best
taken in the context of existing empirical constraints on the models.
Dynamical masses of very young sub-stellar objects did not exist at
the time of the investigation of \citet{neuhauser_etal05}, but the
first dynamical masses were recently reported for the $\sim$1~Myr-old
eclipsing sub-stellar binary 2MASS~J05352184--0546085~A/B
\citep{stassun_etal06}.  We use this young brown-dwarf binary to
decide which set of models is more accurate at predicting the mass of
the objects in this system.  We find that the ``hot-start'' models
reproduce the individual component masses at an age of 1~Myr to within
30\% of their dynamically measured values.
\citeauthor{wuchterl_tscharnuter03}'s evolutionary tracks, on the
other hand, under-predict the dynamical masses by a factor of
$\sim$3---too large a discrepancy to be explained by, e.g., a
potential 1--5~Myr under-estimate of the age of the binary.  While we
acknowledge that 2MASS~J05352184--0546085~A/B provides only two
empirical data points, and that a broader comparison between data and
theory will be needed to conclusively test the models, we consider the
above test a sufficient demonstration that the ``hot-start'' models
are more accurate at ages of 1--3~Myr, and favor their predictions for
the mass of GQ~Lup~B.  Hence, we conclude that GQ~Lup~B is not a
Jupiter-like object in its initial phase of contraction, but rather a
$\geq$10 times heavier brown dwarf.

\section{GQ~LUP: A SUB-STELLAR COMPANION AND A DISK \label{sec_disk}}

\begin{figure}
\plotone{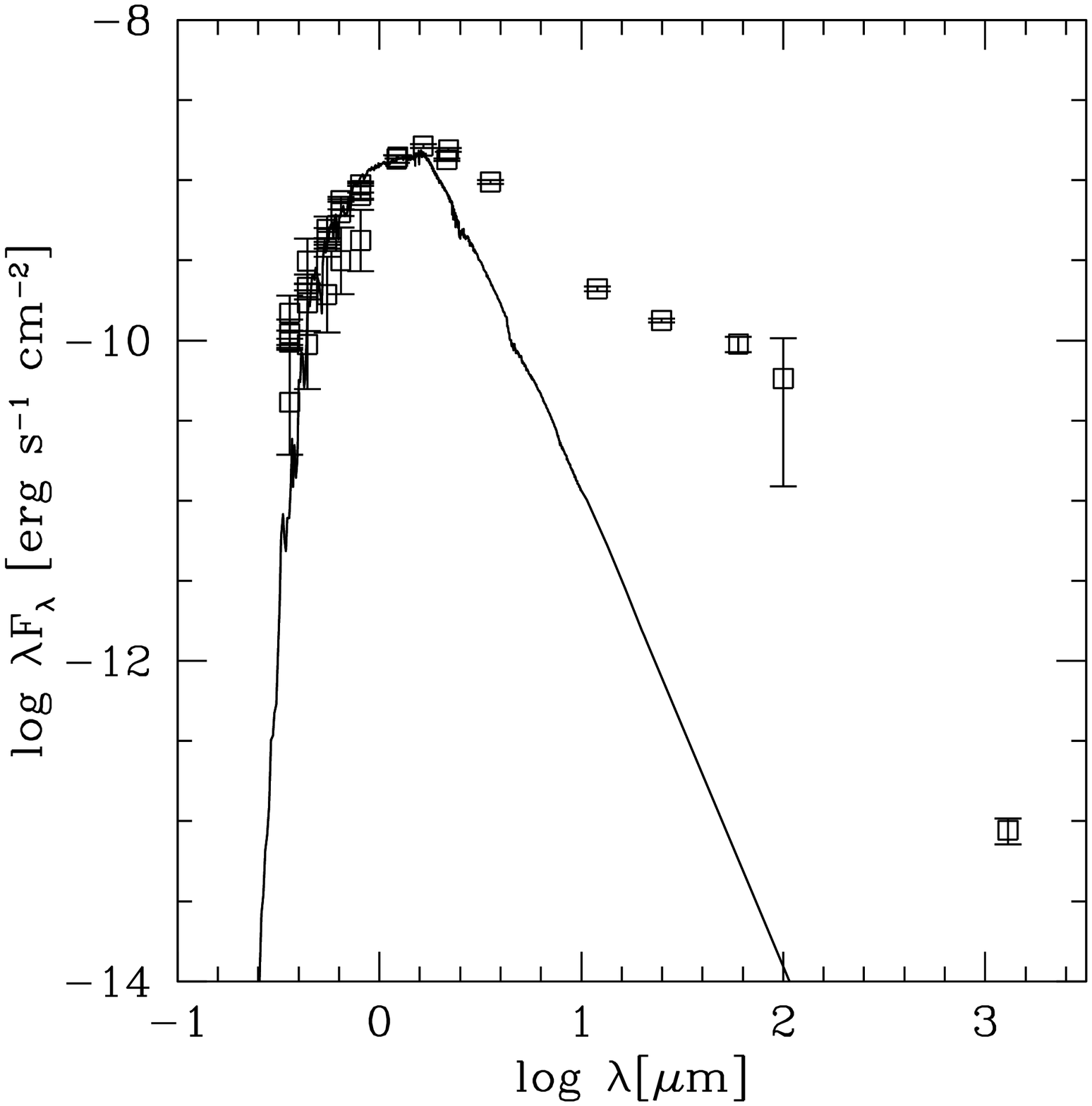} \figcaption{Spectral energy distribution of GQ~Lup.
A 4200~K (spectral type $\approx$M0) NextGen model
\citep{hauschildt_etal99} with solar metallicity and surface gravity
$\log g=4.0$ has been over-plotted.  An extinction of $A_V=1.0$~mag
has been assumed (\S\ref{sec_age}).  The empirical data (squares with
error bars) are from \citet{mundt_bastian80}, \citet{covino_etal92},
\citet{gregoriohetem_etal92}, \citet{hughes_etal94},
\citet{appenzeller_etal83}, unpublished Las Campanas data contained in
William Herbst's T~Tauri star photometry database, 2MASS, {\sl IRAS}
\citep{weaver_jones92}, and \citet{nuernberger_etal97}.  Given the
high optical variability of GQ~Lup (\S\ref{sec_age}) and the
non-simultaneity of the different data sets, we have made no attempt
to fit the data.  Instead, the model photosphere is simply normalized
to the 2MASS flux at 1.2~$\micron$.
\label{fig_sed}}
\end{figure}

GQ~Lup is known to possess a strong IR excess (Fig.~\ref{fig_sed})
from {\sl IRAS} \citep{weaver_jones92} that signals the presence of an
optically thick circumstellar disk.  The co-existence of this disk
with a low-mass sub-stellar companion is highly relevant for theories
of planet and brown-dwarf formation.  The initial claim of
\citet{neuhauser_etal05} that the mass of GQ~Lup~B could be as small
as 1~\Mj\ posed significant difficulties for planet-formation
theories.  The existence of such a young planet at $>$100~AU from its
host star required that the planet formed in the denser inner reaches
($\sim$30~AU) of the stellar system and was then ejected through a
dynamical interaction with a more massive third body \citep{boss06}.
While realizing the obvious bias in favor of discovering such ejected
planets through direct imaging, a conceptually simpler account of the
formation of GQ~Lup~B may offer a more viable solution.  In view of
our present results, the requirement that GQ~Lup~B must have formed
near its present separation can be satisfied given the higher inferred
mass for the companion.  At a minimum mass of 10~$\Mj$, GQ~Lup~B is
more massive than the opacity mass limit for turbulent fragmentation
\citep[$\sim$5~\Mj;][]{bate_etal03}.  Therefore, the GQ~Lup~A/B system
probably formed through direct collapse and fragmentation of the
parent molecular cloud into a binary.  Moreover, as pointed out by
\citet{janson_etal06}, GQ~Lup~A/B is far from unique in the realm of
wide ($>$100~AU) very low mass ratio ($M_2/M_1\lesssim0.03$) binaries,
where it is joined by systems such as HR~7329~A/B
\citep{lowrance_etal00}, AB~Pic A/B \citep{chauvin_etal05b}, and most
recently, HD~203030~A/B \citep{metchev_hillenbrand06} and HN~Peg~A/B
\citep{luhman_etal06}.  The characteristics of these objects suggest
formation through cloud fragmentation, and we infer that GQ~Lup~A/B
followed the same formation scenario.

\section{CONCLUSION}

We have presented near-IR integral field spectroscopic AO observations
of the GQ~Lup~A/B binary system obtained with the OSIRIS IFS on Keck.
Our results demonstrate the utility of adaptive optics IFS
spectroscopy in studying faint close-in sub-stellar companions buried
in the complex speckle-dominated haloes of bright stars.  Our $J$- and
$H$-band spectra of GQ~Lup~B show the typical characteristics of very
young ultra-cool dwarfs, such as the lack of alkali absorption and
triangularly shaped $H$-band continua.  From our $J$-band spectra, we
determine a spectral type of M6--L0 for GQ~Lup~B, in marginal
agreement with previous $K$-band spectroscopy from
\citet{neuhauser_etal05}, who find M9--L4.  We argue that the
difference in the spectral type estimates from the $J$- and $K$-band
spectroscopy arises from the sensitivity to surface gravity of the
2~$\micron$ continuum slope indices used by \citet{neuhauser_etal05},
and that the true spectral type of GQ~Lup~B is indeed earlier.  This
claim is sustained by a comparison of GQ~Lup~B to other sub-stellar
objects of similar luminosities in young stellar associations of
similar ages.  Following a careful analysis of the age and
heliocentric distance of GQ~Lup~A, we conclude that the mass of
GQ~Lup~B is 0.010--0.040~\Msun.  The mass estimate is based on the
``hot-start'' models from \citet{burrows_etal97} and
\citet{chabrier_etal00}, and is $\geq$5--10 times higher than the mass
predicted by the core-accretion---gas-capture models of
\citet{wuchterl_tscharnuter03}.  We favor the use of the ``hot-start''
models because they more accurately predict the dynamical masses of
the newly-discovered \citep{stassun_etal06} 1~Myr brown-dwarf binary
2MASS~J05352184--0546085~A/B.  The inferred mass of GQ~Lup~B makes it
an improbable wide-orbit analog of the present population of
$<$15~\Mj\ radial-velocity extra-solar planets.  Instead, we conclude
that GQ~Lup~A/B is a member of a growing population of wide
($>$100~AU) binary systems with very low mass ratios
($M_2/M_1\lesssim0.03$), the identification of which has only recently
been possible through high-contrast imaging.

\acknowledgements

We thank Wolfgang Brandner, Eric Mamajek, and Michael Meyer for
stimulating and insightful discussions, J.~Davy Kirkpatrick and
Catherine Slesnick for providing us with near-IR spectra of young
brown dwarfs for comparison, and William Herbst for making his
photometry of T Tauri stars available on-line.  The authors would also
like to acknowledge the exceptional efforts of the OSIRIS engineering
team which includes Ted Aliado, George Brims, John Canfield, Thomas
Gasaway, Chris Johnson, Evan Kress, David LaFreniere, Ken Magnone,
Nick Magnone, Juleen Moon, Gunnar Skulason, and Michael Spencer.  We
thank the W.M. Keck Observatory (CARA) staff that were involved in the
installation and commissioning of the OSIRIS instrument.  In
particular, we thank Sean Adkins, Paola Amico, Randy Campbell, Al
Conrad, Allan Honey, Jim Lyke, David Le Mignant, Grant Tolleth, Marcos
Van Dam, and Peter Wizinowich.

We give special thanks to the W.M. Keck Observatory and to the Keck
Science Steering Committee for their invariant support throughout the
development of OSIRIS.  Funds were generously allocated by CARA, the
National Science Foundation (NSF) Telescope System Instrumentation
Program (TSIP), and the NSF Science and Technology Center for Adaptive
Optics (CfAO).  The CfAO funds were managed by the University of
California at Santa Cruz under cooperative agreement No. AST-98-76783.
This publication makes use of data products from the Two Micron All
Sky Survey, which is a joint project of the University of
Massachusettes and the IPAC/California Institute of Technology, funded
by the NASA and the NSF.  Support for S.A.M.\ was provided by NASA
through the {\sl Spitzer} Fellowship Program, under award 1273192.
The authors wish to recognize and acknowledge the very significant
cultural role and reverence that the summit of Mauna Kea has always
had within the indigenous Hawaiian community.  We are most fortunate
to have the opportunity to conduct observations from this mountain.

{\it Facilities:} \facility{Keck (OSIRIS)}


\clearpage

\end{document}